\documentclass[11pt,a4paper]{article}

\usepackage{amsfonts}
\usepackage{amssymb}
\usepackage{graphicx}
\usepackage{amsmath}
\usepackage{color}
\usepackage{enumerate}

\RequirePackage{mathrsfs} \RequirePackage[sc]{mathpazo}
\RequirePackage{wasysym} \RequirePackage{setspace}
\begin{document}

\title{\textbf{Ehrenfest Scheme  of  Higher Dimensional  AdS Black holes in The Third Order Lovelock-Born-Infeld Gravity}}
\author{A. Belhaj$^{1,2}$, M. Chabab$^2$, H. EL Moumni$^2$, K. Masmar$^2$, M.  B. Sedra$^{3}$ \\
{\small $^{1}$D\'epartement de Physique, Facult\'e
Polydisciplinaire, Universit\'e Sultan
 Moulay Slimane, B\'eni Mellal,  Morocco } \\
{\small $^{2}$High Energy Physics and Astrophysics Laboratory, FSSM,
 \small Cadi Ayyad University, Marrakesh, Morocco
} \\
{\small $^{3}$  D\'{e}partement de Physique, LHESIR, Facult\'{e} des
Sciences, Universit\'{e} Ibn Tofail,
 K\'{e}nitra, Morocco.} }
\date{\today } \maketitle


\abstract{
 Interpreting the cosmological constant   as a
thermodynamic pressure and its conjugate quantity as a thermodynamic
volume, we  reconsider the  investigation of  $P$-$V$  critical
behaviors of ($1+n$)-dimensional  AdS  black holes in
Lovelock-Born-Infeld gravity.   In particular, we derive an explicit
expression of the universal number $\chi=\frac{P_c v_c}{T_c}$ in
terms of the space dimension $n$. Then, we examine the phase
transitions at the critical points of such  black holes for $6 \leq
n < 11$ as required  by  the physical condition  of the
thermodynamical quantities  including  criticality behaviors. More
precisely, the Ehrenfest equations have been checked and they reveal
that the black hole system undergoes a second phase transition at
the critical points.}

\newline
\section{\label{sec:level1}Introduction}

Black   holes in various dimensions have received an increasing
attention  in connection with  higher dimensional  supergravity
models  embedded  either in superstrings  or in  M-theory moving on
non trivial  geometric backgrounds  including  Calabi-Yau manifolds.
A particular interest has been devoted   to the study of   extremal
black hole solutions using  the attractor mechanism  developed in
\cite{a,b,c,d}. In this approach, the corresponding effective
potentials and  the entropy functions have been computed  using the
U-duality group theory applied to   the black hole charge
invariants. Models based on Calabi-Yau manifolds
have been elaborated using complex and quaternionic geometries  \cite{d}.\\

 \par Recently,  many  efforts   have been  devoted to  study thermodynamical
properties of  the black  holes using  techniques  explored in
statistical physics  and fluids \cite{30,a1,a2,hasan1,hasan2}. In particular, the
critical behaviors have been obtained for several black holes in
various dimensions using either numerical or analytic calculations
\cite{30,4,5,50,6,7,8,KM}. A special interest has been  on Ads black
holes  in arbitrary dimensions \cite{our}. More precisely, the state
equations $P=P(T,V)$  have been worked out by considering the
cosmological constant as the thermodynamic pressure and its
conjugate as the thermodynamic volume. This activity  has revealed a
nice interplay  between the behavior of the RN-AdS black hole
systems and the Van der Waals fluids which  has been seriously
investigated in many places. In fact,  it has been shown that the
corresponding $P$-$V$ criticality can be related to the liquid-gas
systems of statistical physics. Moreover, it has been seen that  the
criticality depends on the dimension of the spacetime in which   the
black holes live. This subject has been extensively investigated
producing interesting results [16-23].

More recently, a special emphasis has been put on the
thermodynamical properties of   AdS black holes in
Lovelock-Born-Infeld Gravity \cite{TLove}.  A critical behavior in
seven dimensions  has been obtained for uncharged and  charged black
holes. Particularly, the $P$-$V$  diagram has been elaborated for
such black holes   with  spherical geometries.

Motivated by black objects in string theory and the above mentioned
studies, we consider in this work the  criticality of ($1+n)$
dimensional  AdS black holes in Lovelock-Born-Infeld
gravity. Interpreting the cosmological constant as a thermodynamic
pressure and its conjugate quantity as a thermodynamic volume, we
investigate such behaviors in terms of the dimension of the
spacetime and other parameters specified later on. Among others, we
derive an explicit expression of the universal number
$\chi=\frac{P_c v_c}{T_c}$,   for $6 \leq n < 11$ as required  by
the reality of the thermodynamical quantities and criticality. Then,
we discuss the phase transitions at the critical points of these
black hole solutions. In particular, we show that the Ehrenfest
thermodynamical equations are satisfied and find that the black hole
system undergoes a second phase transition.

\section{Thermodynamics of  higher dimensional  black holes in  the third order Lovelock-Born-Infeld gravity }
In this section, we focus on the study of thermodynamics of
 higher dimensional  black holes in Lovelock-Born-Infeld gravity. In
particular, we obtain an explicit expression for the corresponding
state equation at critical points in $(1+n)$ dimensions.  The
analysis will be made in  terms of  three parameters: the space
dimension $n$, the Born-Infeld parameter $\beta$ and the curvature
constant $k$.  The discussion of the  critical behaviors will be
given in  next sections.

To start, we consider the physical action describing  the third
order Lovelock gravity  in the presence of a nonlinear Born-Infeld
electromagnetic gauge  field as studied in \cite{x3,ref1,ref3}. This
action, which has been investigated in different context,  takes the
following form
\begin{equation}
I_G=\frac{1}{16 \pi}\int
d^{n+1}x\sqrt{-g}(-2\Lambda+\mathcal{L}_1+\alpha_2 \mathcal{L}_2
+\alpha_3 \mathcal{L}_3+L(F)),
\end{equation}
where $\Lambda$ is the cosmological constant.  $\mathcal{L}_1$,
$\mathcal{L}_2$, $\mathcal{L}_3$ and $L(F)$ represent
Einstein-Hilbert, Gauss-bonnet, the third order Lovelock and  the
Born-Infeld Lagrangians respectively. They are given by  the
following, expressions
\begin{eqnarray}
\mathcal{L}_1&=& R,\\
\mathcal{L}_2&=& R_{\mu\nu\gamma\delta}R^{\mu\nu\gamma\delta}-4R_{\mu\nu}R^{\mu\nu}+R^2,\\
\mathcal{L}_3&=& 2 R^{\mu\nu\sigma\kappa}R_{\sigma\kappa\rho\tau}R^{\rho\tau}_{\mu\nu} +8R^{\mu\nu}_{\sigma\rho}R^{\sigma\kappa}_{\nu\tau}R^{\rho\tau}_{\mu\kappa}+24 R^{\mu\nu\sigma\kappa}R_{\sigma\kappa\nu\rho}R^\rho_\mu\\ \nonumber
&+& 3R R^{\mu\nu\sigma\kappa}R_{\sigma\kappa\mu\nu}+24 R^{\mu\nu\sigma\kappa}R_{\sigma\mu}R_{\kappa\nu}+16 R^{\mu\nu}R_{\nu\sigma}R^\sigma_\mu-12 R R^{\mu\nu}R_{\mu\nu}+R^3,\\
L(F)&=&4\beta^2\left(1-\sqrt{1+\frac{F^2}{2\beta}} \right).
\end{eqnarray}
where  the $\beta$ is the Born-Infeld parameter as proposed in
\cite{ref1,ref3}. The constants $\alpha_{1,2}$ read as follows
\begin{equation}
\alpha_2=\frac{\alpha}{(n-2)(n-3)}
\end{equation}
\begin{equation}
\alpha_3=\frac{\alpha^2}{72    \left(\begin{array}{c}n-2 \\4\end{array}\right) },
\end{equation}
 where $\alpha$ is  the  Lovelock coupling constant. This choice has been made for simplicity reason. It's used due to  some properties  which   are absent  in the Gauss- Bonnet gravity with the  three fundamental constants. Other choices can be made as repported in \cite{ref3}.
 
  The $(n + 1)$-dimensional static solution
is given by
 \begin{equation}
 ds^2=-f(r)dt^2+\frac{dr^2}{f(r)}+r^2d\Omega^2_{n-1}.
 \end{equation}
In this solution,  the black hole function $f(r)$ takes the
following form
 \begin{equation}
 f(r)=1+\frac{r^2}{\alpha}(1-g(r)^{1/3})
 \end{equation}
 where
 \begin{equation}
 g(r)=1+\frac{3 \alpha m}{r^n}-\frac{12
  \alpha
  \beta^2}{n(n-1)}[1-\sqrt{1+\eta}-\frac{\Lambda}{2\beta^2}+\frac{(n-1)\eta}{(n-2)}F(\eta)].
 \end{equation}
 $F(\eta)$ is the hypergeometric function
\begin{equation}
F(\eta)=_2F_1\left(
[\frac{1}{2},\frac{n-2}{2n-2}],[\frac{3n-4}{2n-2}],-\eta \right)
\end{equation}
with
\begin{equation}
\eta=\frac{(n-1)(n-2)q^2}{2\beta^2r^{2n-2}}
\end{equation}
  It is interesting to  note  that  the $(n-1)$ dimensional line element  $d\Omega^2_{n-1}$ depends on the geometry in
  question and it is given by
\begin{equation}
d\Omega^2_{n-1} =
d\theta_1^2\sum\limits_{i=2}^{n-1}\prod\limits_{j=1}^{i-1}sin^2\theta_j
  d\theta_i^2
\end{equation}
defining $(n-1)$ dimensional hypersurfaces with the  constant
curvature $(n-1)(n-2)$. For  these configurations, the Hawking
temperature  is expressed  as  in  \cite{x3}

\begin{equation}
\label{temp}
T=\frac{(n-1)[3(n-2)r_+^4+3(n-4)\alpha r_+^2+(n-6)\alpha^2]+12r_+^6\beta^2(1-\sqrt{1+\eta})-6\Lambda r_+^6}{12 \pi(n-1)r_+(r_+^2+\alpha)^2}
\end{equation}

Similar calculations  reveal that  the  entropy reads as
\begin{equation}
\label{entropy}
S=\int_0^{r_+}\frac{1}{T}\left(\frac{\partial m}{\partial r_+}\right)=\frac{\Sigma_k(n-1)r_+^{n-5}}{4}\left (
 \frac{r_+^4}{n-1}+\frac{2r_+^2\alpha}{n-3}+\frac{\alpha^2}{n-5} \right
 ).
\end{equation}
It follows that this  physical solution requires that the integer
$n$  is  constrained by the condition:  $n\geq 6$. It is also
remarked
 that, for higher dimensional cases,  the Lovelock gravity does not  coincide with
 Einstein one. However,   the usual four dimensional  case can be recovered by deleting
the Lovelock gravity  terms \cite{Zou:2013owa}.

It is recalled that   that $\Sigma_k$ describes the volume of the
$(n-1)$ dimensional hypersurface and the thermodynamic volume can be
written as
\begin{equation}
\label{4} V=\frac{\Sigma_k r_+^n}{n}.
\end{equation}
 It is known  that the Gibbs free energy is given by
\begin{eqnarray}\hspace{-2cm}\nonumber
G&=& \frac{\Sigma_k r_+^{n-6}}{48 \pi\alpha (r_+^2+ \alpha)^2}[
(n-1)r_+^6(r_+^2+\alpha)^2\left(-1+\frac{(r_+^2+\alpha)^3}{r_+^6}+\frac{12\alpha\beta^2\left(1-\frac{\Lambda}{2\beta^2}-\sqrt{\eta+1}+\frac{(n-1)\eta F(\eta)}{n-2}\right)}{n(n-1)}\right)
\\\nonumber
&-&\alpha\left(   \frac{r_+^4}{n-1}+\frac{2  r_+^2 \alpha}{n-3}+\frac{\alpha^2}{n-5} \right)\left((n-1) \left(3(n-2)r_+^4+3(n-4)k\alpha r_+^2+(n-6)\alpha^2\right)\right.\\
&-&\left.6\Lambda r_+^6+12 r_+^6 \beta^2(1-\sqrt{\eta+1})\right)]
\end{eqnarray}
 We have
considered only the internal energy  to discuss $P-V$ criticality in
the extended phase space. In fact, it has been realized that the
internal energy has been needed to give a complete study on the
corresponding criticality.

Since  the thermodynamical pressure of the black hole is interpreted as  the  cosmological
constant,
\begin{equation}
\label{p2} P=-\frac{\Lambda}{8\pi},
\end{equation}
  the first law of the  black hole thermodynamics can be modified  by the
introduction of the variation of the cosmological contant (the
pressure) in this law. Now, one   can write the following mass
equation
\begin{equation}
dm= T dS+ \Phi dq+PdV+ \mathcal{B}d\beta
\end{equation}
where $\Phi$ is the electric potential and    $\mathcal{B}$ is a
conjugate quantity  to $\beta$ called Born-Infeld vacuum
polarization \cite{7, KM,Zou:2013owa}. This quantity is given by
\begin{eqnarray}\nonumber
\mathcal{B}&=&\left(\frac{\partial m}{\partial \beta }\right)_{S,q,P,\beta}=\frac{\Sigma_k}{8\pi n \beta r_+^2}\left\{  2 r_+^{2n}\beta^2 \left(2-\sqrt{4+\frac{2(n-1)(n-2)q^2r_+^{2(1-n)}}{\beta^2}}\right)           \right.\\
&+& \left. (n-2)(n-1)q^2r_+^2 \; _2F_1\left(     \left[\frac{1}{2},\frac{n-2}{2n-2}\right] ,\left[\frac{n-2}{2n-2}\right],-\frac{(n-1)(n-2)q^2}{2\beta^2 r^{2(n-1)}}       \right)\right\}.
\end{eqnarray}

In fact, we  can obtain the pressure as a function of the
temperature and the horizon radius. Indeed, after a lengthy but
straightforward calculations, we  show that  the the pressure reads
as

\begin{eqnarray}\nonumber
P&=& \frac{T}{v}%
-\frac{ ((n-2) (n-1) v-32 \pi  \alpha  T)}{\pi  (n-1)^2 v^3}%
-\frac{16 \alpha   ((n-4) (n-1) v-16 \pi  \alpha  T)}{\pi  (n-1)^4 v^5}%
-\frac{256 \alpha ^2  (n-6)}{3 \pi  (n-1)^5 v^6}\\ &+& \frac{\beta ^2 \left(\sqrt{\frac{2^{4 n+1} (n-2) (n-1)^3 q^2 v^2 ((n-1) v)^{-2
   n}}{\beta ^2}+64}-8\right)}{32 \pi }.
\end{eqnarray}

By confronting this equation to the Van der Waals equation of state,
we readily derive the specific volume as follows
\begin{equation}
v=\frac{4r_+}{n-1}.
\end{equation}

Hereafter,  we focus our analysis on critical behaviors and   we
show how  to establish an explicit expression of the  universal
number $\chi=\frac{P_c v_c}{T_c}$ in $(1+n)$ dimensions. Then, we
study the phase diagram transitions using  the classical
thermodynamical physics.

\section{Critical behavior description}
As mentioned  before,  here we consider the study of   the critical
behaviors of the above black hole solutions. We first  give a detail
study on uncharged case. Then we  shortly present  the charged case.
\subsection{Uncharged solutions}
 Roughly
speaking,  the computation leads to  the following state equation
\begin{eqnarray}\nonumber
P&=&  \frac{(n-1)^2 v^2 (\pi  (n-1) T v- (n-2))-16 \alpha   ( (n-4)-2 \pi  (n-1) T v)}{\pi  (n-1)^3 v^4}\\
&-&\frac{256 \alpha ^2  (k (n-6)-3 \pi  (n-1) T v)}{3 \pi
(n-1)^5 v^6}.
\end{eqnarray}
In fact,  through numerical calculations, we plot the  $P$-$V$
 diagrams  in terms of the space dimension $n$. The results are shown in  figure 1.

\begin{center}
\begin{figure}[!ht]
\begin{tabbing}
\hspace{9cm}\=\kill
\includegraphics[scale=0.63]{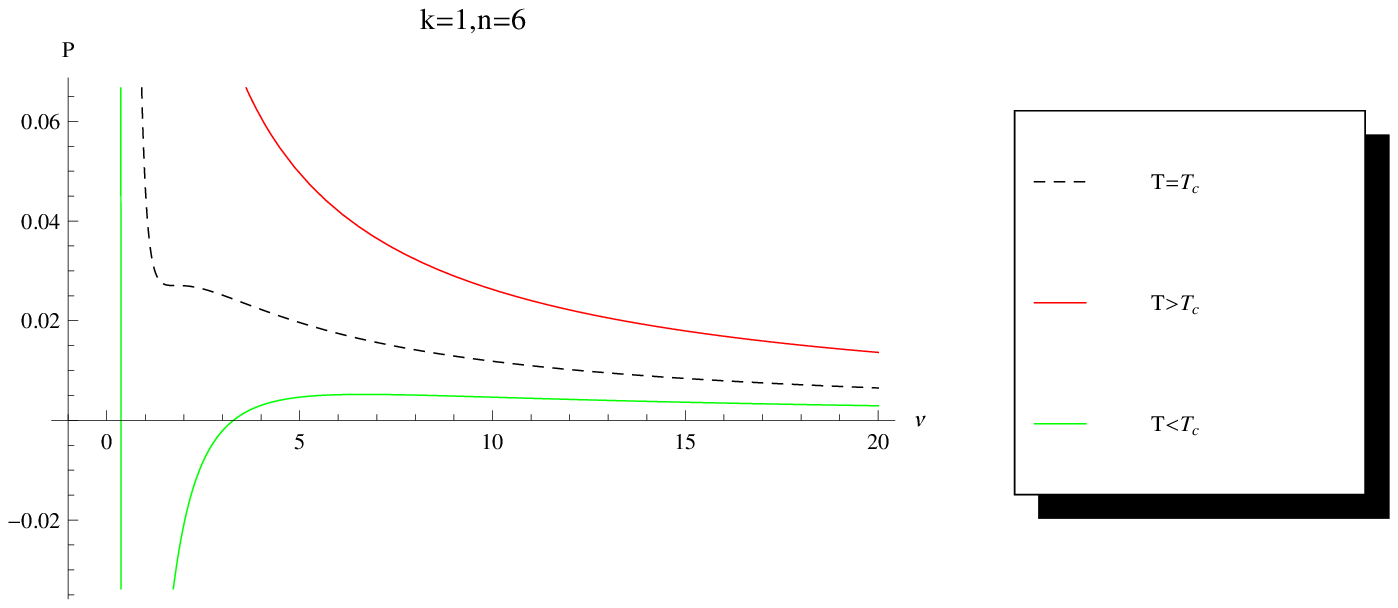}\>\includegraphics[scale=0.63]{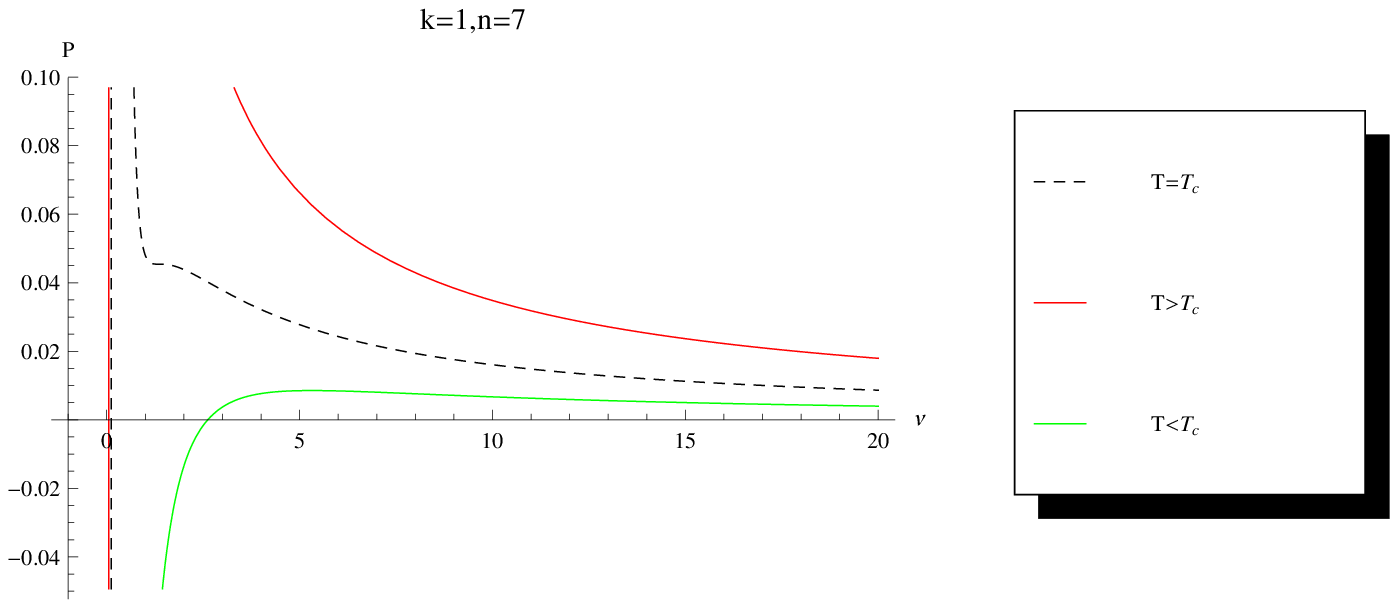} \\
\includegraphics[scale=0.63]{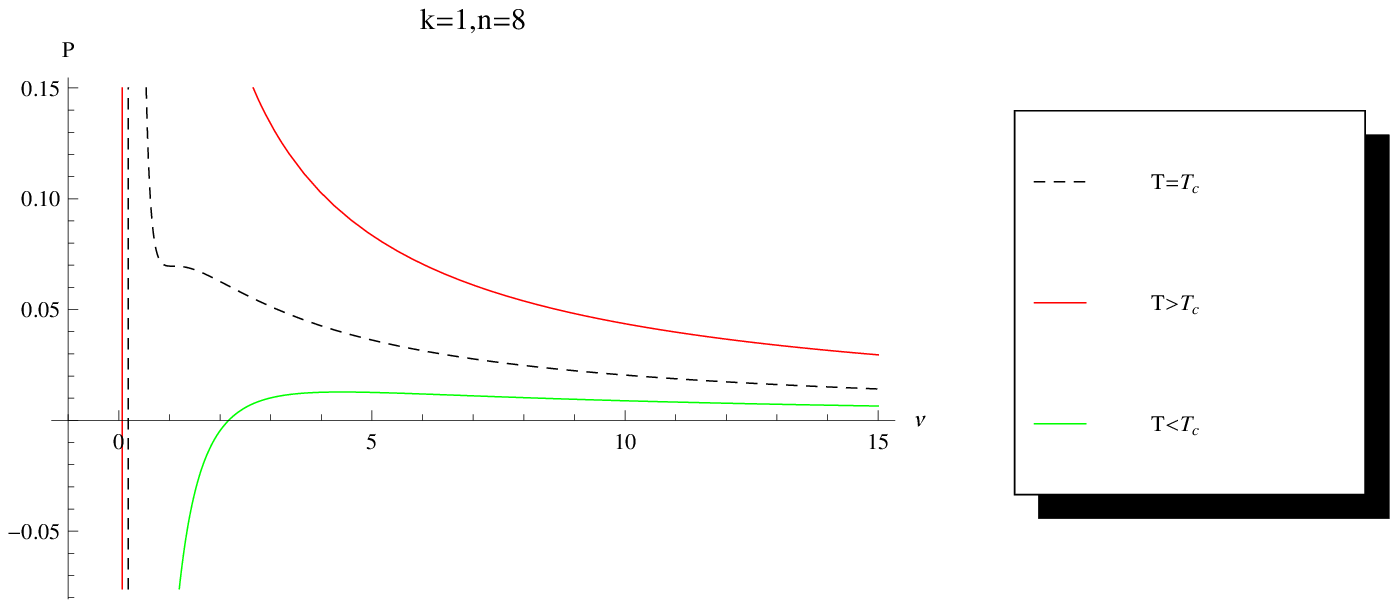}\> \includegraphics[scale=0.63]{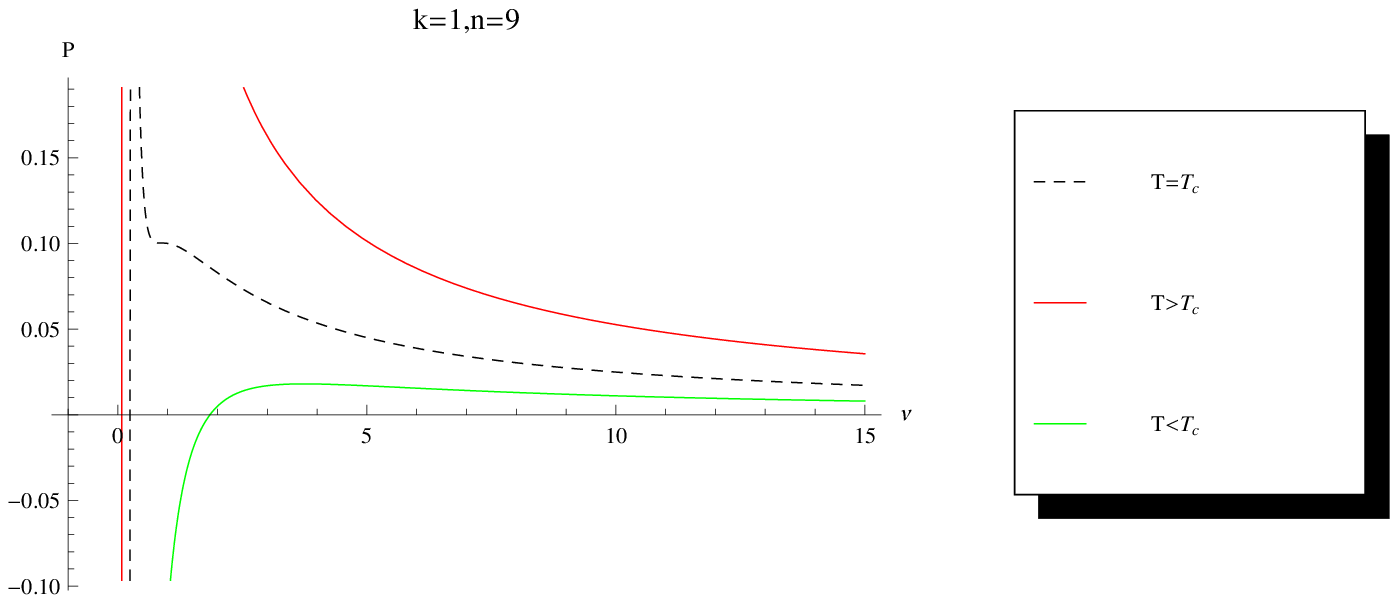}\\
\includegraphics[scale=0.63]{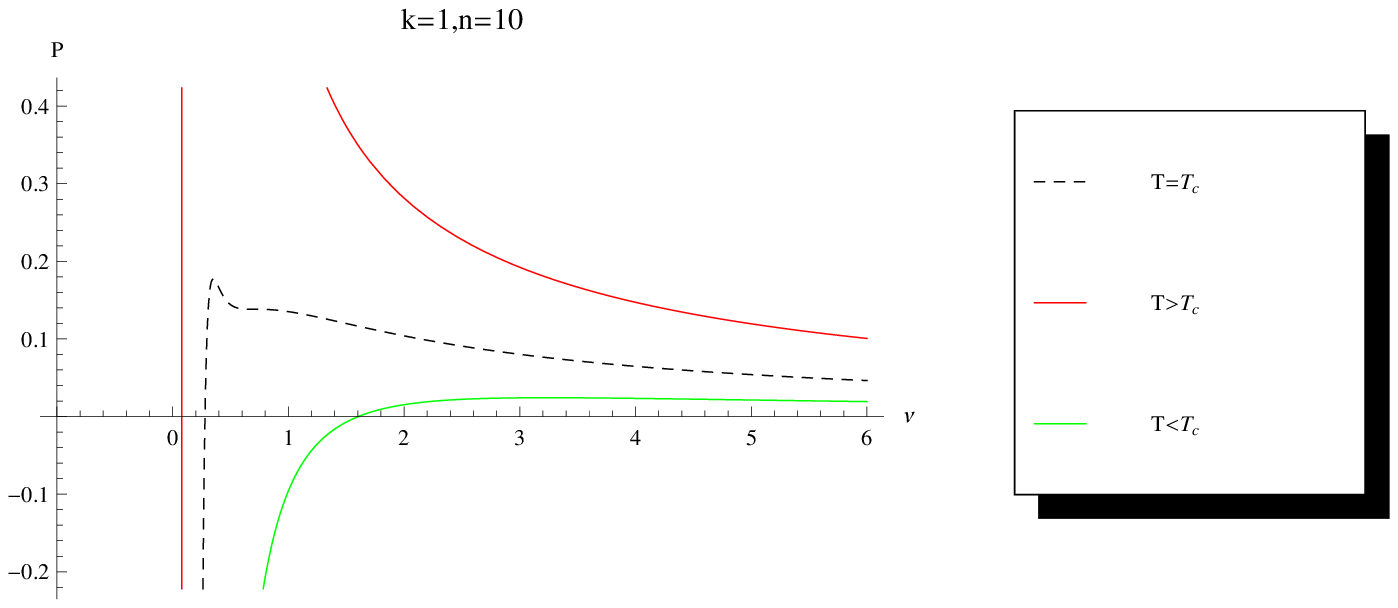}\>\includegraphics[scale=0.63]{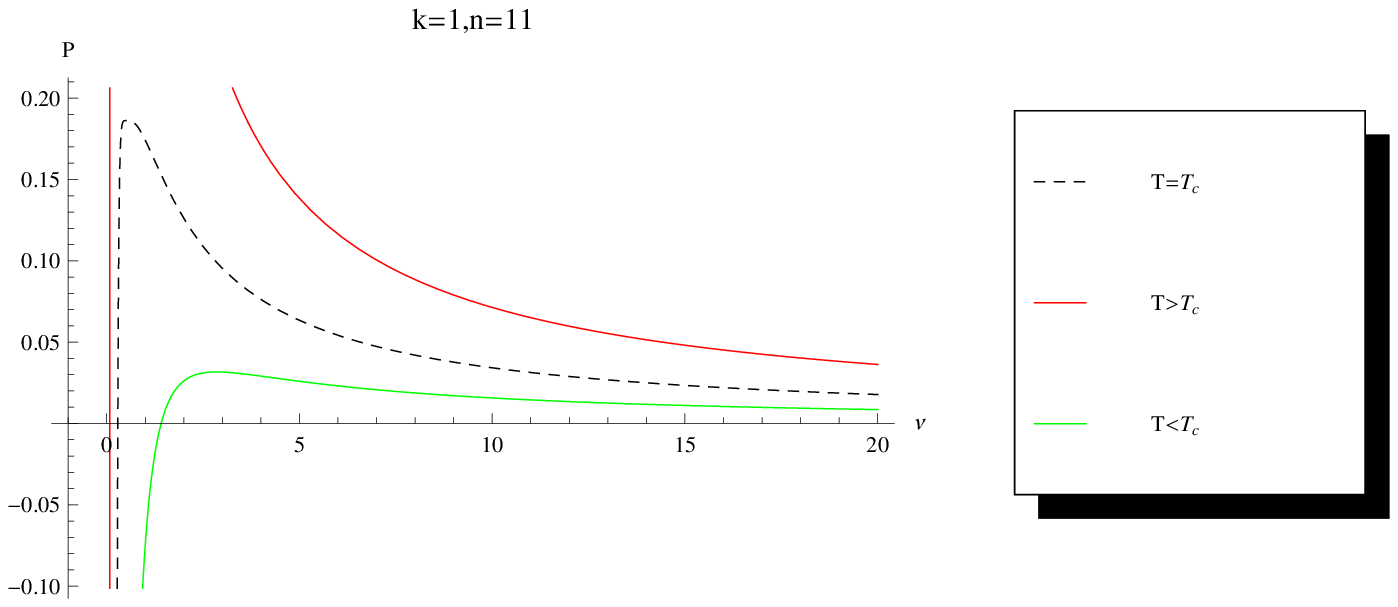} \\
\end{tabbing}
\vspace*{-.2cm} \caption{The $P-V$ diagrams  for space dimension $n$ between $6$ and $11$.
$T_c$ is the critical temperature for  $\alpha=1$.}
\label{fig1}
\end{figure}
\end{center}

From this figure, we  observe  that the  $P$-$V$ behavior is similar
to the Van der Waals'one. This   may allow to derive the critical
point coordinates. To do this, we  should first solve the following
system of equations
\begin{equation}\label{system}
\frac{\partial P}{\partial v}=0, \quad \frac{\partial^2 P}{\partial v^2}=0.
\end{equation}
Then, as a consequence, one can determine the explicit  thermodynamical
expressions for  the  critical values.   They are given by,

\begin{eqnarray}\label{criticx}\nonumber
T_c&=& -\frac{ \left((n-1)^2 (n (2 n-29)+2)-\sqrt{(11-n)
(n-1)^5} (n+14)\right)}{2 \pi  (n-2) (n-1)^7 \left(\frac{(n+4)
(n-1)^2+2 \sqrt{(11-n)
   (n-1)^5}}{(n-2) (n-1)^4}\right)^{5/2} \sqrt{\alpha }}\\
v_c&=4&  \sqrt{\frac{(n+4) (n-1)^2+2 \sqrt{(11-n)
(n-1)^5}}{(n-2) (n-1)^4}} \sqrt{\alpha }\\\nonumber
 P_c&=&\frac{(n-2)^2 (n-1)^9
C(n)}{48 \pi  \left((n+4) (n-1)^2+2 \sqrt{(11-n) (n-1)^5}\right)^5
\alpha }.
\end{eqnarray}

where $C(n)$ is a quantity depending on the space dimension $n$
which can be  expressed as

\begin{eqnarray}
 C(n)&=&(n (n (n (9 n-536)+4772)+3328)-6048) (n-1)^2\\&+&
\nonumber  2 \sqrt{(11-n) (n-1)^5} (n ((68-11 n) n+1420)-432).
\end{eqnarray}

By combining the critical  expressions shown in Eq. (\ref{criticx}),  we can
deduce the explicit form of  the universal number $\chi=\frac{P_c
v_c}{T_c}$. Indeed, this number is given by
\begin{equation}\small
\chi= -\frac{ (n-1)^4 C(n)}{6 \left((n+4) (n-1)^2+2 \sqrt{(11-n) (n-1)^5}\right)^2 D(n)}
\end{equation}
where   the quantity  $D(n)$ reads as
\begin{equation}
 D(n)=(n-1)^2 (n (2 n-29)+2)-\sqrt{(11-n) (n-1)^5} (n+14).
\end{equation}

It is worth noting the system of equations (\ref{system}) involves
actually
 two real solutions. However,  we exclude the critical one which yields a vanishing
 critical specific volume, hence $r_{+c}=0$, for $n=6$  producing   a black hole without
 event horizon, known as nude singularity. Besides, unlike the excluded solution,
  the other solution reproduces exactly the critical coordinates  derived in \cite{x3}.

Here we stress that the $\chi$ generalized expression involves many
interesting features.  First,  we  recover the  six dimension result
given in \cite{x3}. Indeed, for $n=6$, the critical coordinates are
given by
\begin{equation}
T_c=\frac{1}{\pi\sqrt{5\alpha}},\quad v_c=\frac{4\sqrt{\alpha}}{\sqrt{5}},\quad P_c=\frac{17}{200\pi\alpha},\quad \chi=\frac{P_c v_c}{T_c}=\frac{17}{50}.
\end{equation}
Furthermore, the universal number $\chi$ behaves nicely in terms of
the space dimension as illustrated in figure 2. Thus,  we observe
from figure 1 that  critical behaviors with a clear inflexion point
appear  only when the space dimension lies in the range $6\leq n <
11$.

For $n=11$,   it follows   that  for a temperature less than the
critical one,  the behavior of the black hole does not show an
inflexion point but a maximum. However, the latter can not be
considered this point like a critical one. For this reason, we have
considered only  models associated with $n< 11$.

\begin{center}
\begin{figure}[!ht]
\begin{center}
\includegraphics[scale=1]{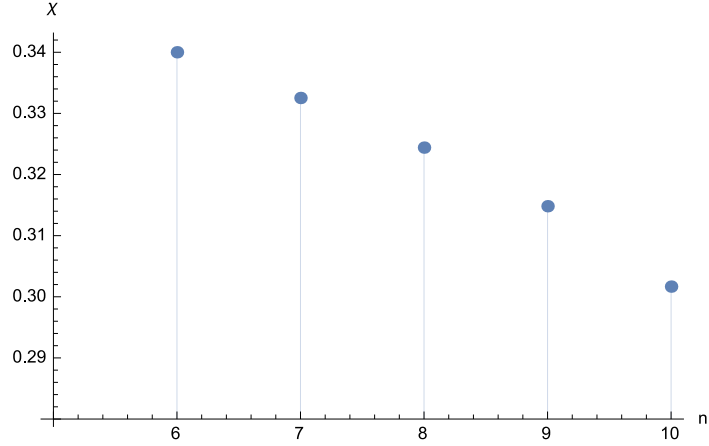}
\end{center}
 \caption{The universal number in term of space-time dimension.} \label{fig2} \vspace*{-.2cm}
\end{figure}
\end{center}

It is noted  that $n$ do not exceed  space dimension $10$ as
required by the physical condition  of the critical volume. This is
not a surprising feature  in  high energy physics. In fact,  a close
inspection in higher dimensional theories shows that this critical
dimension appears naturally in string theory and related topics.
Indeed, $n=10$  corresponds to  a  non perturbative limit of eleven
dimensional type IIB superstring. This limit is interpreted in terms
of $12$  dimensional theory. As proposed by  Vafa,    this is known
as  F-theory   which  has been constructed using  a  geometric
interpretation of the  $SL(2,Z)$ duality \cite{vafa}. This
observation motivates  us to think about a string theory realization
of these black holes in terms of the brane
physics. We hope to come back to this issue in future works. \\

\subsection{Charged solutions}

Here,   we shortly give  the calculations for  the charged black
holes in the asymptotic limit of $\beta$ ($\beta\rightarrow\infty$).
In this limit,  the spherical  topologies show critical behaviors
for $6 \leq n < 11$. The numerical evaluations are listed
hereinafter which agree with \cite{1zz}. In particular, we
illustrate graphically the critical behaviors. We plot in figure 3
the $P$-$V$ diagrams in various dimensions.
\begin{table}[!ht]
{\center
 \begin{tabular}{|c||c||c||c||c||c|}
 \hline
 q&$\alpha$&$T_c$&$v_c$&$P_c$&$\frac{P_cv_c}{T_c}$  \\ \hline
0.5&1& 0.1855  &  1.3731 &  0.0452 &  0.3349   \\ \hline 2&1& 0.1839
&  1.4728 &  0.0438 &   0.3509   \\ \hline
 1&1&  0.1851  &  1.4036 & 0.0448 &  0.3401    \\ \hline
 1&0.5& 0.2537 &  1.1748& 0.0807  &    0.3737    \\ \hline
   1&2& 0.1313 &  1.9254 & 0.0226  &    0.3328     \\ \hline

 \end{tabular}
 \caption{Critical values for $   n = 7, \beta\rightarrow\infty$}}
\end{table}

\begin{table}[!ht]{\center
 \begin{tabular}{|c||c||c||c||c||c|}
 \hline
 q&$\alpha$&$T_c$&$v_c$&$P_c$&$\frac{P_cv_c}{T_c}$  \\ \hline
0.5&1& 0.2292  &  1.1308 & 0.0684  & 0.3374  \\ \hline 2&1&  0.2287
&  1.1561 & 0.0677  &   0.3426   \\ \hline
 1&1&  0.2297  & 1.1071  & 0.0689  &   0.3323    \\ \hline
 1&0.5& 0.3157 & 0.9437  & 0.1243  &  0.3716      \\ \hline
   1&2& 0.1627 & 1.5188  & 0.0347  &    0.3245     \\ \hline

 \end{tabular}
 \caption{Critical values for $   n = 8, \beta\rightarrow\infty$}}
\end{table}

\begin{table}[!ht]{\center
 \begin{tabular}{|c||c||c||c||c||c|}
 \hline
 q&$\alpha$&$T_c$&$v_c$&$P_c$&$\frac{P_cv_c}{T_c}$  \\ \hline
0.5&1& 0.2747  &  0.9182 &  0.0987 & 0.3298   \\ \hline 2&1& 0.2742
&  0.9389 &  0.0979 &   0.3352   \\ \hline
 1&1& 0.2751  &  0.8990 & 0.0993  &   0.3245    \\ \hline
 1&0.5& 0.3786 & 0.7844  &  0.1786 &    0.3701    \\ \hline
   1&2& 0.1949 &  1.2253 & 0.0500  &    0.3149     \\ \hline

 \end{tabular}
 \caption{Critical values for $   n = 9, \beta\rightarrow\infty$}}
\end{table}

\begin{table}[!ht]{\center
 \begin{tabular}{|c||c||c||c||c||c|}
 \hline
 q&$\alpha$&$T_c$&$v_c$&$P_c$&$\frac{P_cv_c}{T_c}$  \\ \hline
0.5&1& 0.3210 & 0.7631 &  0.1358 &   0.3229 \\ \hline 2&1&  0.3205 &
0.7816  & 0.1348  &   0.3287   \\ \hline
 1&1&  0.3214  & 0.7453  & 0.1367  &   0.3170    \\ \hline
 1&0.5& 0.4422 & 0.6688  &  0.2439 &  0.3689      \\ \hline
   1&2& 0.2277 &  0.9944 &  0.0691 &   0.3018      \\ \hline

 \end{tabular}
 \caption{Critical values for $   n = 10, \beta\rightarrow\infty$}}
\end{table}

\vspace{1 cm}

\begin{center}
\begin{figure}[!ht]
\begin{tabbing}
\hspace{9cm}\=\kill
\includegraphics[scale=0.83]{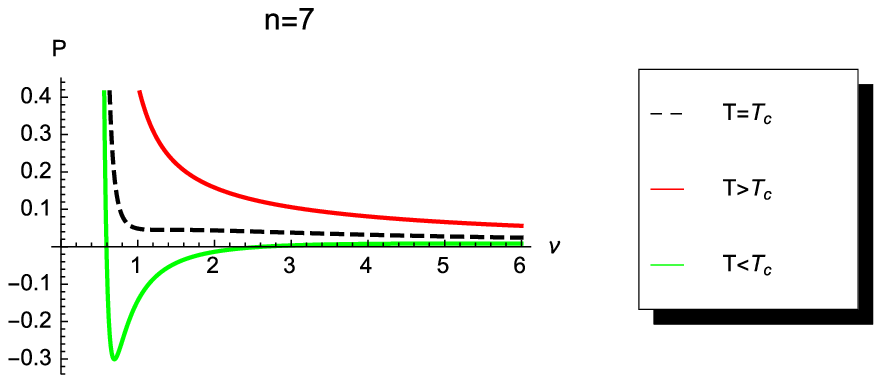}\>\includegraphics[scale=0.83]{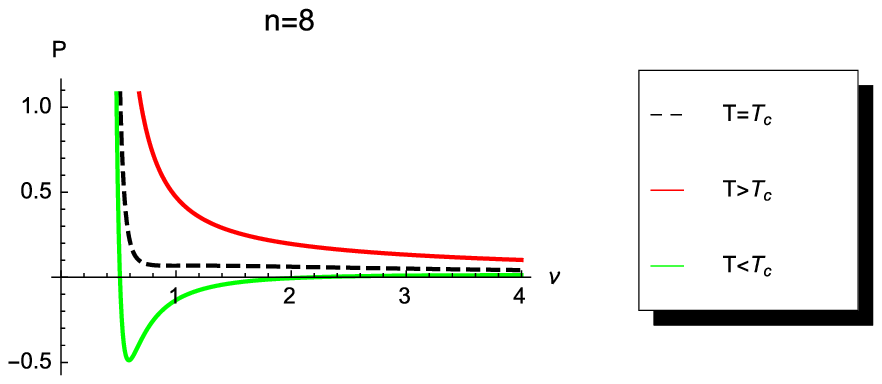} \\
\includegraphics[scale=0.83]{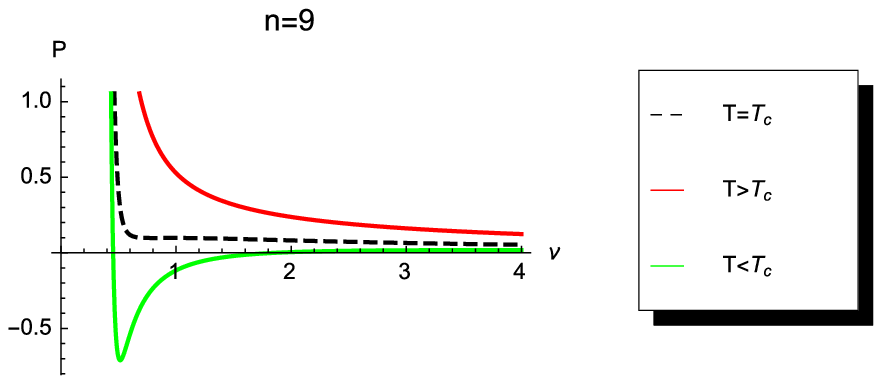}\> \includegraphics[scale=0.83]{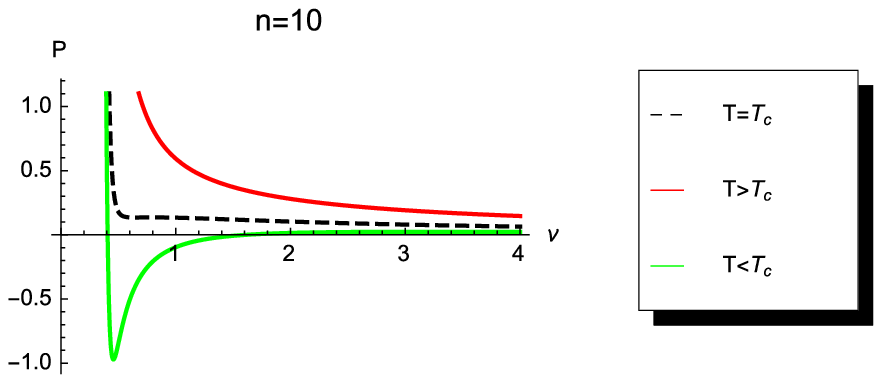}\\
\end{tabbing}
\vspace*{-.2cm} \caption{The $P-V$ diagrams  for space dimension $n$
between $6$ and $11$. $T_c$ is the critical temperature for
$\alpha=1$,  $q=0.5$ and $\beta\rightarrow\infty$.} \label{fig3}
\end{figure}
\end{center}
 To make contact with charged case, we  discuss the corresponding critical behavior. This has been presented
  in   figure
 4.  In particular, we  plot   the equation of state with the Born-infeld  parameter $\beta=1$
and make   comparison between the Born-Infeld theory and the Maxwell
one in  the limit where $\beta\rightarrow\infty$.

\begin{center}
\begin{figure}[!ht]
\begin{tabbing}
\hspace{9cm}\=\kill
\includegraphics[scale=0.83]{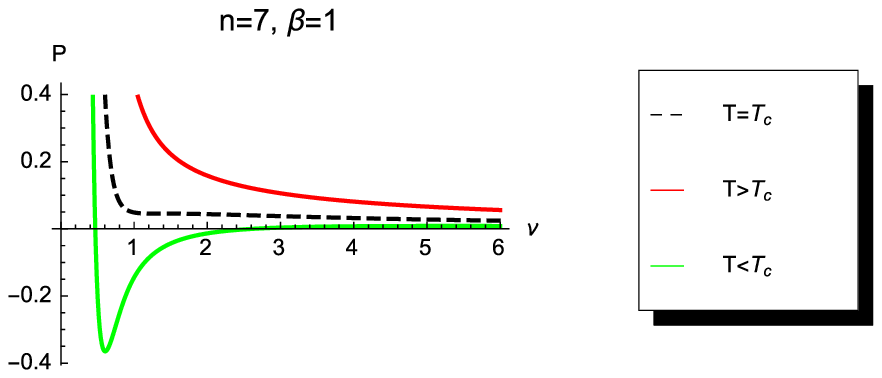}\>\includegraphics[scale=0.83]{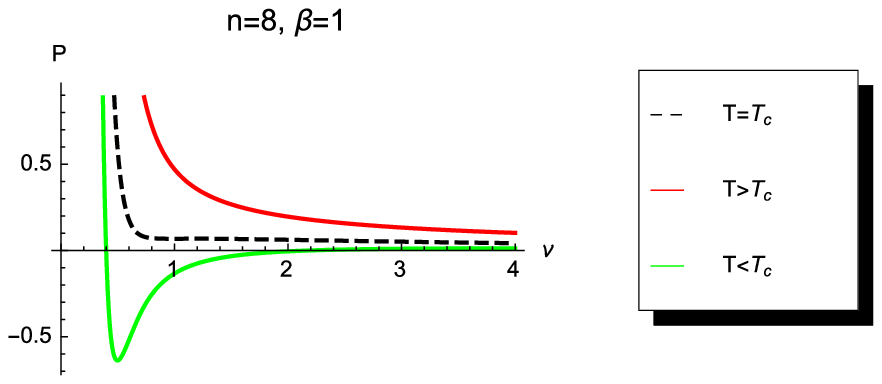} \\
\includegraphics[scale=0.83]{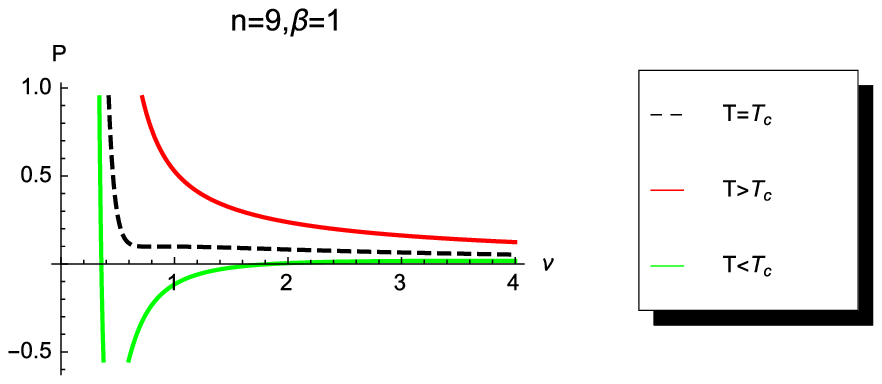}\> \includegraphics[scale=0.83]{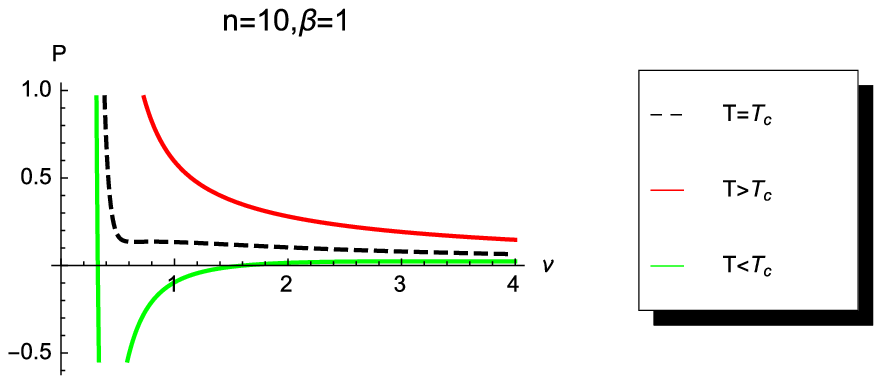}\\
\end{tabbing}
\vspace*{-.2cm} \caption{The $P-V$ diagrams  for space dimension $n$
between $6$ and $11$. $T_c$ is the critical temperature for
$\alpha=1$ and $q=0.5$. } \label{fig4}
\end{figure}
\end{center}

It is observed form  figure $4$ that the Born-Infeld parameter
modify the critical points in the  the $(P,V)$ plan. It has been
shown that  $P_{c}({\beta\rightarrow\infty})> P_{c}({\forall
\beta})$.

\section{Ehrenfest scheme}
Having discussed the $P$-$V$  criticality of  $(1+n)$ dimensional
 AdS black holes in Lovelock-Born-Infeld Gravity, we move
now to the study  of the corresponding phase transitions using
classical thermodynamics principals. We note that the classification
of such phases associated with  the  first order and higher orders
can be done in terms of the Clausius-Clapeyron-Ehrenfest equations.
Indeed, the first order transition  is ensured  when  the
Clausius-Clapeyron equation is satisfied at the critical points.
However,  the second order transition arises when the Ehrenfest
thermodynamical equations are verified. In this section, we examine
such  equations using results obtained in classical thermodynamics
\cite{x30,x31,x32}. In fact,  the Ehrenfest equations read as
\begin{equation}\label{x13}
 \left(\frac{\partial P}{\partial T}
  \right )_S= \frac{C_{P_2}-C_{P_1}}{VT(\Theta_2-\Theta_1)}=\frac{\Delta C_P}{VT\Delta
  \Theta},
 \end{equation}
  \begin{equation}\label{x14}
 \left(\frac{\partial P}{\partial T} \right )_V= \frac{\Theta_2-\Theta_1}
 {\kappa_{T_2}-\kappa_{T_1}}=\frac{\Delta \Theta}{\Delta
 \kappa_{T}}.
 \end{equation}
In these equations,  $\Theta=\frac{1}{V}\left(\frac{\partial
V}{\partial T}\right)_P$ is  the volume expansion and
$\kappa_T=\frac{1}{V}\left(\frac{\partial V}{\partial P}\right)_T$
defines  the isothermal compressibility coefficient. \\

In what follows, we compute the relevant thermodynamical quantities
involved in the above equations for ($1+n$) dimensional
AdS black holes in Lovelock-Born-Infeld gravity. Indeed, combining
equations (\ref{temp}),  (\ref{entropy})  and  (\ref{p2}), we  get a
general expression of the temperature

\begin{equation}\label{xtemp}
T=\frac{ (n-1) \left( (n-6) \alpha ^2+3  (n-4) \alpha  \zeta
(S)^2+3 (n-2) \zeta (S)^4\right)+48 \pi  P \zeta (S)^6}{12 \pi
(n-1) \zeta (S) \left(
   \alpha +\zeta (S)^2\right)^2}
\end{equation}

where $\zeta(S)$ is the real positive root solving  the entropy
function equation (\ref{entropy}).   Performing similar
calculations, we can also determine the specific heat at constant
pressure and the volume expansion coefficient. They  are
respectively given by

\begin{eqnarray}\label{cp}
C_P&=&\frac{\zeta (S) \left( \alpha +\zeta (S)^2\right) \left(
(n-1) \left( (n-6) \alpha ^2+3  (n-4) \alpha  \zeta (S)^2+3
(n-2) \zeta (S)^4\right)+48 \pi
   P \zeta (S)^6\right)}{B(\alpha ,S,P)\zeta '(S)}\nonumber\\
\label{teta} \Theta &=&\frac{12 \pi  (n-1) n \zeta (S) \left(
\alpha +\zeta (S)^2\right)^3}{B(\alpha ,S,P)}
\end{eqnarray}

 where   the function $B(\alpha ,S,P)$ reads as

\begin{eqnarray}
\nonumber B(\alpha ,S,P)&=&- (n-6) (n-1) \alpha ^3-2  (n-9) (n-1) \alpha ^2 \zeta (S)^2+48 \pi  P
   \zeta (S)^8\\
&& +18  (n-1) \alpha  \zeta (S)^4-3  \zeta (S)^6 ((n-3) n-80 \pi  P \alpha +2)
  \end{eqnarray}
Thanks to the famous  thermodynamic relation
\begin{equation}
\left(\frac{\partial V}{\partial P}\right)_T \left(\frac{\partial
P}{\partial T}\right)_V \left(\frac{\partial T}{\partial
V}\right)_P=-1,
\end{equation}
we obtain  the expression of  the  isothermal compressibility
coefficient
\begin{equation}\label{kt}
K_T=\frac{48 \pi  n \zeta (S)^6 \left( \alpha +\zeta
(S)^2\right)}{B(\alpha ,S,P)}.
\end{equation}

From  these equations,  we notice  the existence of a  special
factor appearing in  their dominators. This factor can be explored
to stress critical behaviors for  the above thermodynamic
quantities. In fact, the constraint  $B(\alpha ,S,P)=0$ leading to a
divergence of the heat capacity can easily be checked for the
critical points.

To discuss the   validity of Ehrenfest equations at the critical
points, we  should  analyze the expression of    the  volume
expansion coefficient $\Theta$. The latter is evaluated as
\begin{equation}
 V\Theta=\left( \frac{\partial V}{\partial T} \right)_P=\left(
 \frac{\partial V}{\partial S} \right)_P \left( \frac{\partial S}{\partial T}
 \right)_P=\left( \frac{\partial V}{\partial S} \right)_P\left( \frac{C_P}{T}
 \right).
\end{equation}
Moreover, the right handed side  of Eq. $(\ref{x13})$ can be
converted to
\begin{equation}
 \label{RHS}
 \frac{\Delta C_P}{T V \Delta \Theta}=\left[\left(\frac{\partial S}{\partial V}\right)_P
 \right]_c,
 \end{equation}
where the  index  $c$ indicates the values of  the thermodynamical
variables  at the critical points. Exploring  Eqs.
$(\ref{entropy})$, $(\ref{4})$ and $(\ref{RHS})$, we  obtain
  \begin{equation}
  \label{A}
 \frac{\Delta C_P}{T V \Delta \Theta}=\left[
 \frac{\pi ^{\frac{1}{2} (-n-1)} \Gamma \left(\frac{n+1}{2}\right) \zeta (S)^{1-n}}{2 \zeta '(S)}
 \right]_c.
 \end{equation}
 Using  Eq. $(\ref{xtemp})$, the left handed side  of Eq. $(\ref{x13})$
 translates to
  \begin{equation}
  \label{B}
 \left[\left(\frac{\partial P}{\partial T}\right)_S \right]_c=\frac{(n-1) \left( \alpha +\zeta
  (S_c)^2\right)^2}{4 \zeta (S_c)^5}.
 \end{equation}
Similar calculations can be done using
 Eqs.(\ref{entropy}), (\ref{4}) and (\ref{xtemp}). In this way,  the left handed side  of Eq.$(\ref{x14})$
 becomes
\begin{equation}
\label{xxA} \left[\left(\frac{\partial P}{\partial T}\right)_V
\right]_c=\frac{(n-1) \left( \alpha +\zeta (S_c)^2\right)^2}{4
\zeta (S_c)^5}.
\end{equation}
A close inspection of the expressions of the  isothermal
compressibility coefficient $K_T$ and volume expansion coefficient
$\Theta$ shows that we have
\begin{equation}
 V K_T=-\left( \frac{\partial V}{\partial P} \right)_T=\left( \frac{\partial T}{\partial P}
  \right)_V \left( \frac{\partial V}{\partial T} \right)_P=\left( \frac{\partial T}{\partial P} \right)_V V
  \Theta.
\end{equation}
The right handed side  of Eq. $(\ref{x14})$  produces the following
formula
\begin{equation}
\label{xxB}
 \frac{\Delta \Theta}{\Delta K_T}=\left[\left(\frac{\partial P}{\partial T}\right)_V \right]_c=\frac{(n-1) \left( \alpha +\zeta (S_c)^2\right)^2}{4 \zeta (S_c)^5}
 \end{equation}
revealing the validity of the second Ehrenfest's equation.

It is worth to note that the Prigogine-Defay (PD) ratio
\cite{defay, defay1} can also be computed.  Indeed, the calculation shows the following expression,
 \begin{equation}\label{PD}
 \Pi =\frac{\Delta C_P \Delta K_T}{T V (\Delta \Theta)^2}=\left[ \frac{\pi ^{\frac{1}{2}
  (-n-1)} \Gamma \left(\frac{n-1}{2}\right) \zeta (S)^{6-n}}{\zeta '(S) \left( \alpha
   +\zeta (S)^2\right)^2} \right]_c.
 \end{equation}

To illustrate the above analysis,  we consider  the case of $n=7$
associated with an eight  dimensional black solution. In this case,
the expression of the $\zeta$  is reduced  to
\begin{equation}
 \zeta(S)=\sqrt{\frac{\sqrt[3]{\pi^4  \alpha ^3+12 S}}{\pi ^{4/3}}-
 \alpha
 }.
\end{equation}
Moreover, Eqs. (\ref{A}) and (\ref{B})  become
\begin{eqnarray}
\label{xA}\nonumber
  \frac{\Delta C_P}{T V \Delta \Theta}&=&\left[\left(\frac{\partial
   S}{\partial V}\right)_P \right]_c=\left[\left(\frac{\partial P}
   {\partial T}\right)_S \right]_c\\ &=&\frac{3 \left(\pi ^5  \alpha^3+12
    \pi  S_c\right)^{2/3}}{2 \left(\sqrt[3]{\pi ^4  \alpha ^3+12 S_c}-\pi ^{4/3}  \alpha
    \right)^{5/2}}.
\end{eqnarray}
indicating clearly the  validity of  Ehrenfest  first equation at the critical
point.  Furthermore Eqs. $(\ref{xxA})$ and $(\ref{xxB})$ give the relation
\begin{equation}
\label{xxxB}
 \frac{\Delta \Theta}{\Delta K_T}=
 \left[\left(\frac{\partial P}{\partial T}\right)_V \right]_c=\frac{3 \left(\pi ^5
 \alpha ^3+12 \pi  S_c\right)^{2/3}}{2 \left(\sqrt[3]{\pi ^4  \alpha ^3+12 S_c}-\pi ^{4/3}
   \alpha \right)^{5/2}}.
\end{equation}
It follows that the  Ehrenfest  second equation  is also  valid at
the critical point.

\begin{center}
\begin{figure}[!ht]
\begin{tabbing}
\hspace{9cm}\=\kill
\includegraphics[scale=0.60]{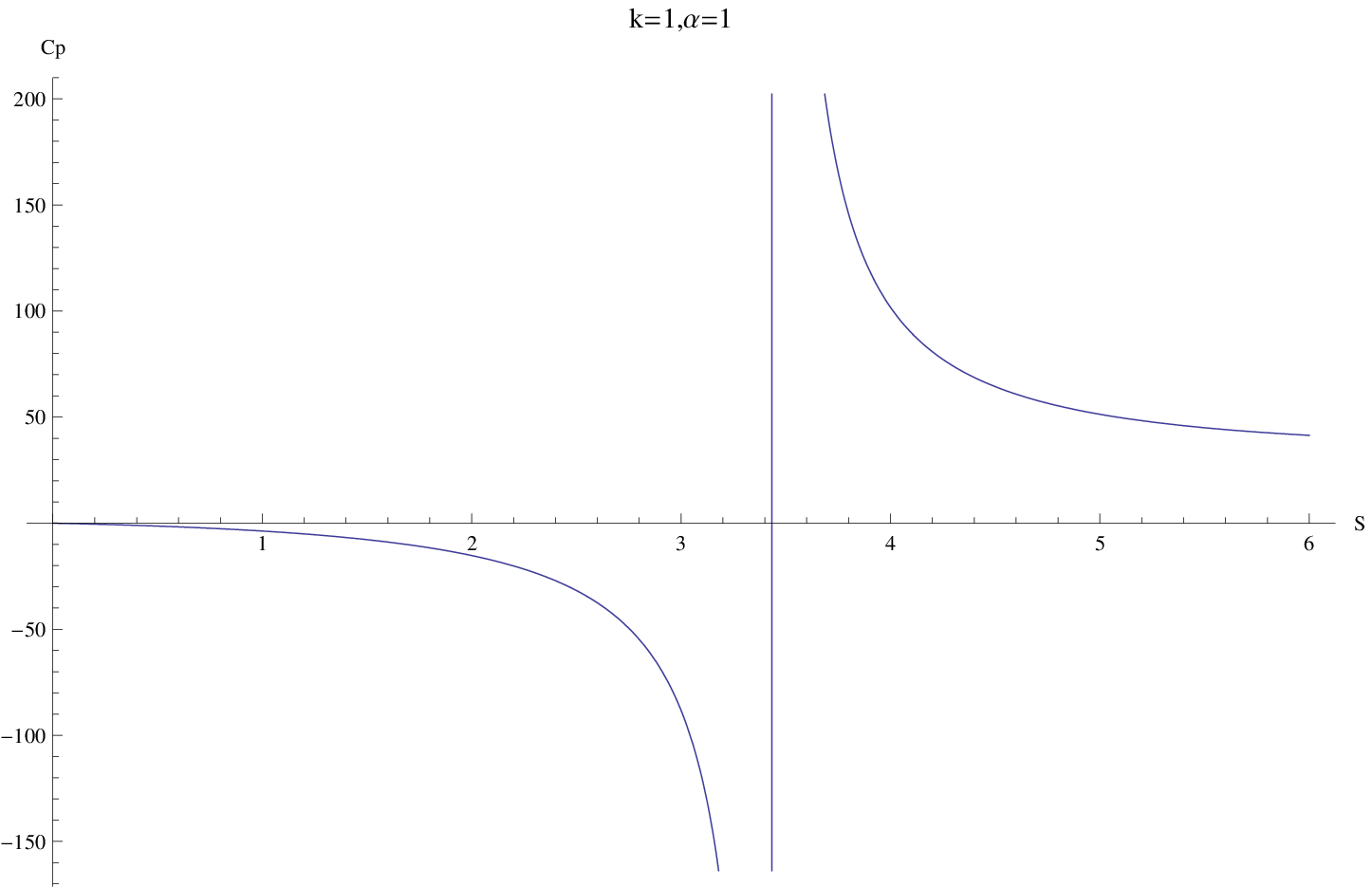}\\\includegraphics[scale=0.60]{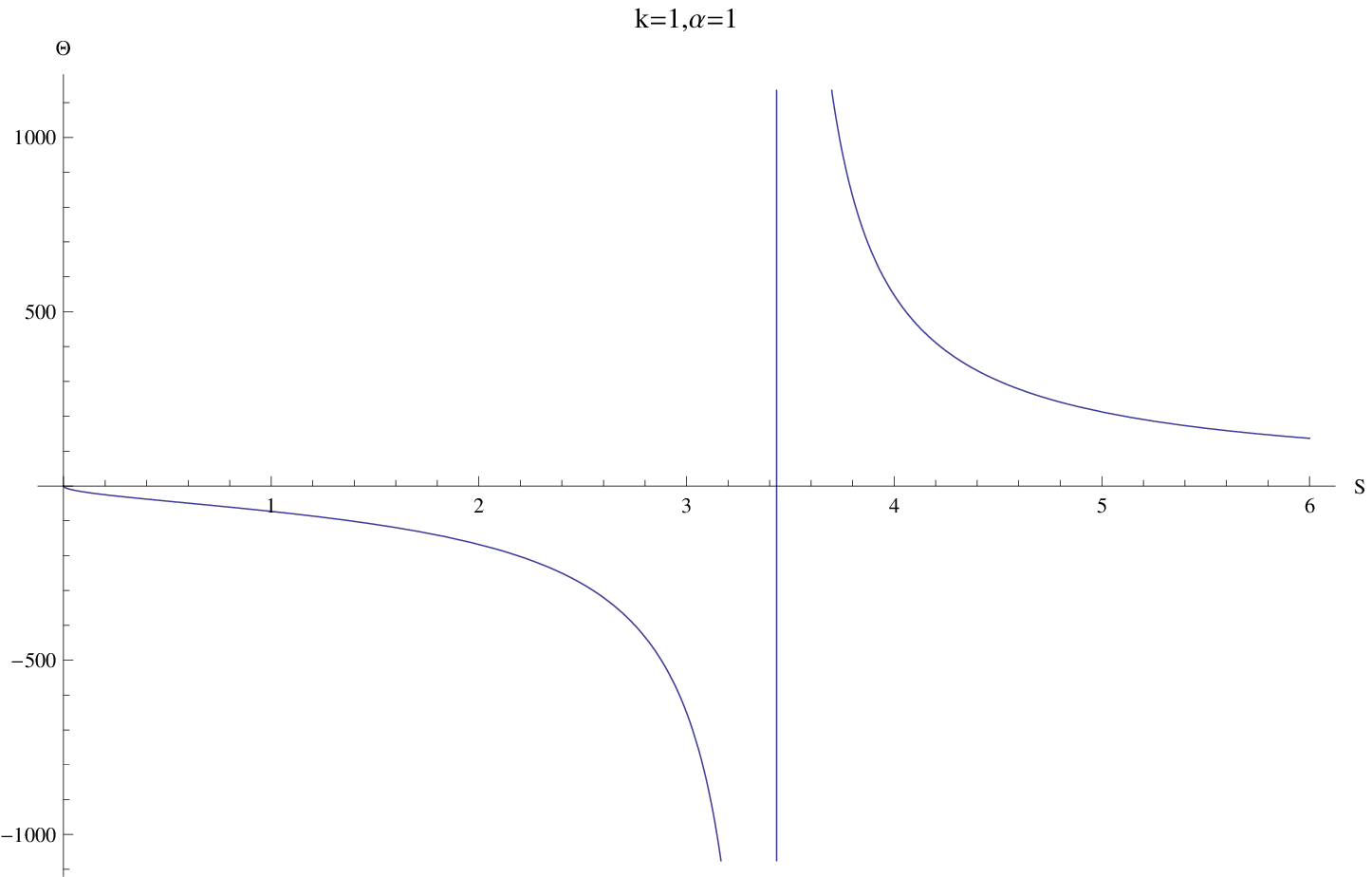}\\\includegraphics[scale=0.60]{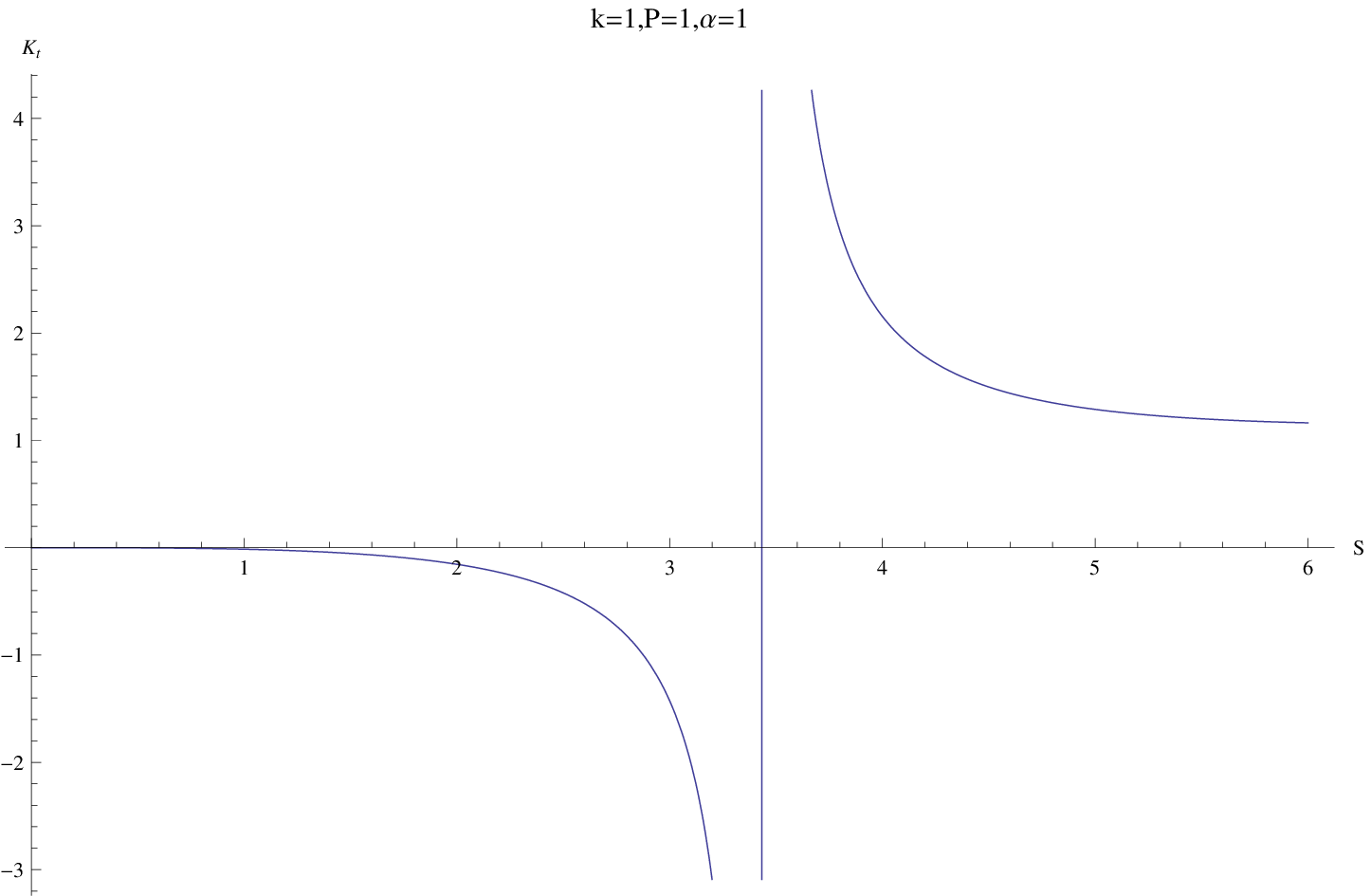} \\
\end{tabbing}
\vspace*{-.2cm} \caption{ $C_p$, $\Theta$ and $k_T$ in terms of the
entropy for $\alpha=1$.} \label{fig5}
\end{figure}
\end{center}

It is worth noting  that  the definition of PD ratio was proposed by
Prigogine and Defay \cite{defay} and reviewed in many works
including  \cite{defay1}. The second Ehrenfest equation is not
always satisfied and the PD ratio can be used to measure the
deviation from the second Ehrenfest equation \cite{xxxx}. Here, here
it reduces to
 \begin{equation}
 \Pi =1,
 \end{equation}
 showing  a  phase transition despite existence of the divergency near
the critical point. Note that   this  matches perfectly
with  the  second order equilibrium transition discussed in
\cite{zzz,zzz1}. For more detail on this case, we plot  all
quantities  in figure \ref{fig5} to illustrate that  such quantities
are divergent at the critical points.


\section{Conclusion and open questions}

In this paper, we have reconsidered  criticality
of  $(n+1)$ dimensional   AdS  black holes in
Lovelock-Born-Infeld gravity. Interpreting the cosmological constant
as a thermodynamic pressure and its conjugate quantity as a
thermodynamic volume, we  have studied thermodynamical behaviours in
terms of  the space dimension $n$.   More  precisely, we have derived
an explicit expression of the universal number $\chi=\frac{P_c
v_c}{T_c}$ in terms of $n$. Then, we have  discussed the phase
transitions at the critical points. In particular, the Ehrenfest
thermodynamical equations have been verified showing that the black
hole system undergoes a second phase transition.

This work comes  up with many open  questions. An  interesting one
concerns the   space dimension $n$. It has been realized  that  the
integer $n$  is constrained to lie within the range
 $$6 \leq n  < 11$$
  as required with the reality of
  the  values of   the physical quantities at the critical  points. A
 fast inspection shows that dimensions between 6 and 11 appear naturally in the study
 of higher dimensional theories including superstrings and  M-theory. This observation  may
 provide a
new challenge on such black holes and theirs  connections   with
string theory compactification.  We believe that the above range can
be explored to investigate possible  realizations in terms of brane
physics. This  issue deserves a more deeper study which could be
addressed in coming works .

\end{document}